\documentclass[useAMS,usenatbib,amsmath,amssymb,floatfix]{mn2e}
\usepackage{amssymb}
\usepackage{graphicx}
\usepackage{bm}
\newcommand{\ve}{\bmath}
\newcommand{\m}{\mathbfss}

\title[Non-parametric reconstruction of the primordial power spectrum at horizon scales from WMAP data]{Non-parametric reconstruction of the primordial power spectrum at horizon scales from WMAP data}
\author[Domenico Tocchini-Valentini, Yehuda Hoffman and Joseph Silk]{Domenico Tocchini-Valentini$^{1}\,^{2}$\thanks{E-mail: dtv@astro.ox.ac.uk, dtv@pha.jhu.edu}, Yehuda Hoffman$^{3}$\thanks{E-mail: hoffman@huji.ac.il}
and Joseph Silk$^{1}$\thanks{E-mail: silk@astro.ox.ac.uk}\\
$^{1}$Department of Physics, University of Oxford,
Denys Wilkinson Building,
Keble Road, Oxford OX1 3RH, United Kingdom\\
$^{2}$Department of Physics and Astronomy, The Johns Hopkins University, Baltimore, MD 21218, USA\\
$^{3}$Racah Institute of Physics, Hebrew University, Jerusalem 91904, Israel}
\begin{document}

\date{Accepted ... Received ...; in original form ...}

\pagerange{\pageref{firstpage}--\pageref{lastpage}} \pubyear{2005}

\maketitle

\label{firstpage}

\begin{abstract}
We extend to large scales a method proposed in previous work that
reconstructs non-parametrically the primordial power spectrum from cosmic
microwave background data at high resolution. The improvement is necessary to
account for the non-gaussianity of the \emph{Wilkinson Microwave Anisotropy
Probe} (WMAP) likelihood due primarily to cosmic variance. We assume the
concordance $\Lambda$CDM cosmology, utilise a smoothing prior and perform
Monte Carlo simulations around an initial power spectrum that is scale-free
and with spectral index $n_{s}=0.97$, very close to the concordance
spectrum. The horizon scale for the model we are considering corresponds to
the wavenumber $k_{h}=4.52\times 10^{-4} \, \mathrm{Mpc^{-1}}$. We find some
evidence for the presence of features and we quantify the probabilities of
exceeding the observed deviations in WMAP data with respect to the fiducial
models. We detect the following marginal departures from a scale-free
(spectral index $n_{s}=0.97$) initial spectrum: a cut-off at $0.0001<k<0.001
\, \mathrm{Mpc^{-1}}$ at 79.5\% (92\%), a dip at $0.001<k<0.003 \,
\mathrm{Mpc^{-1}}$ at 87.2\% (98\%) and a bump at
$0.003<k<0.004 \, \mathrm{Mpc^{-1}}$ at 90.3\% (55.5\%) confidence level.

These frequentist confidence levels are calculated by integrating over the
distribution of the Monte Carlo reconstructions built around the fiducial
models. 
The frequentist analysis finds the low $k$ cutoff of the estimated
power spectrum to be  about $2.5\sigma$ away from the $n_s=0.97$ model, while
in the Bayesian analysis the model is about $1.5\sigma$ away from the estimated
spectrum. (The $\sigma$'s are different for the two different methods.)

\end{abstract}
%\date{\today}
%\maketitle


\section{Introduction}
The spectacular results provided by the Wilkinson Microwave Anisotropy Probe (WMAP) experiment (Bennett et al. 2003) offer the possibility of testing the nature of the initial conditions that seeded the structure of the Universe. The inflationary paradigm puts forward a convincing picture that explains the formation of the first seeds. However inflation has not been yet incorporated successfully in the framework of fundamental physics. Therefore hints from observations may prove useful in discerning such an important connection. 

An increasing number of studies have 
 been devoted to the initial power spectrum reconstruction, due to its potential importance. For example in \cite{sper,peiris,bridle,barger}; Mukherjee \& Wang (2003); Mukherjee \& Wang (2005) a shape parametrized a priori is used. Model independent methods that use so-called non parametric algorithms have also been proposed e.g. in \cite{gaw,tegzald,matsu1,matsu2,kogo1,kogo2,kogo3,shafi,hu,hanne,dtv}; Leach (2005); Sealfon et al. (2005). 
In \cite{dtv} we proposed a method based on regularized least squares to
 non-parametrically reconstruct  under a general smoothing assumption the primordial power spectrum from CMB temperature data in the wave number range $0.01<k<0.1 \, h \, \mathrm{Mpc^{-1}}$, with $h$ standing for the Hubble constant in units of $100 \, \mathrm{Km \, sec^{-1} \, Mpc^{-1}}$. Here we extend the analysis to large scales by proposing the necessary modifications dictated by the non-gaussianity of the data errors caused by cosmic variance. While we fix the late-time cosmological parameters to the best fit values obtained from the combined WMAP and Sloan Digital Sky Survey (SDSS) data (Tegmark et al. 2004), our method has the virtue of allowing an estimate of the power spectrum, at the highest resolution to date, that permits a refined local search for deviations from a scale free power spectrum. The horizon scale for the model we are considering translates into the wavenumber $k_{h}\simeq2\pi/\eta_{0}=4.52\times 10^{-4} \, \mathrm{Mpc^{-1}}$, where $\eta_{0}$ is the conformal time at present.
 A high resolution reconstruction at large scales (using the Richardson-Lucy
 algorithm) has been presented previously by \cite{shafi}, that used binned
 data after the hexadecapole. We propose here a different deconvolution
method that  starts from the unbinned WMAP data and provides for the first time an estimate of the errors in the reconstruction. Deconvolving WMAP data we find indications for the following three features that deviate from a fiducial scale free (spectral index $n_{s}=0.97$) initial spectrum: a cut-off at $0.0001<k<0.001\, \mathrm{Mpc^{-1}}$ at 79.5\% (92\%), a dip at $0.001<k<0.003 \,
\mathrm{Mpc^{-1}}$ at 87.2\% (98\%) and a bump at
$0.003<k<0.004 \, \mathrm{Mpc^{-1}}$ at 90.3\% (55.5\%).
Similar features were also detected in \cite{shafi} and suggestions of a similar cutoff were also provided by \cite{wang1,bridle,contaldi,cline}. The cutoff effect is due mainly to the low large scale power at the quadrupole and to a minor extent to the octopole present in WMAP data (Bennett et al. 2003) and, since it is intriguingly present also in the COBE DMR data (Bennett et al. 1996), certainly deserves some consideration. 

We wish to emphasize that our approach is very similar in spirit to traditional image reconstruction methods and that our aim is \emph{not} to search for the minimal parametric description of the power spectrum. Given a dense enough discretization to closely approximate the integral in Eq.(\ref{integral}) below, the data automatically select the degree of smoothing necessary to allow the highest resolution in the reconstructed signal compatible with the actual noise level, being the result of a tradeoff between resolution and noise filtering. Strictly speaking we are using a large number of band amplitudes as parameters, however since they correspond to a dense grid in wavenumber space that translates in a good approximation of the continuum, we call our method \emph{non-parametric}. As will become apparent later, our method is capable of estimating simultaneously the large number of amplitude parameters. Widely used multi-parameter estimation methods such as Monte Carlo Markov Chain would, in their simplest implementations, find some difficulties in evaluating hundreds of parameters. After having tested the reliability of the method through Monte Carlo simulations, the reconstruction can be compared with simple parametric shapes to look for local deviations, that we estimated with two types of errors, as detailed below.

In Section II we introduce in detail the method; in Section III we show the
outcome of some tests on mock data that allow us to proceed with confidence
to the reconstruction from the actual WMAP temperature data, our main
result. We conclude with a d iscussion of our results in Section IV. 

\section{Estimation of the Power Spectrum}

\subsection{Introduction to the Method}
\label{subsec:intro}
The problem at hand is how to estimate the primordial power spectrum $P(k)$ from a CMB observational data base which is compressed into the form of the angular power spectrum, ${ C_l}$. 

We start first with a pedagogical example where the cosmic variance is neglected, a situation applicable in the large $l$ case.

The temperature CMB angular power spectrum is related to the
primordial power spectrum, $P(k)$, by a convolution with a window
function that depends on the cosmological parameters:
\begin{equation} \label{integral}
C_{\ell}=4\pi\int\Delta^{2}_{\ell}(k)\, P^0(k) \, \frac{\mathrm{d} k}{k} \simeq
\sum_{i}W_{\ell i}\,x_{i},
\end{equation}
where the last term represents a numerical
approximation to the integral obtained through a fine discretization and $x_{i}\equiv P(k_{i})/k_{i}$. We fix the units in the transfer function in order to express the primordial fluctuation spectrum $x$, corresponding to the comoving curvature perturbation spectrum $\Delta_{\mathcal{R}}^{2}$ in \cite{verde}, in units of $2.95\times 10^{-9}$.

Let us suppose that our scope is to estimate a solution for the linear problem
\begin{equation}
\ve{d}=\m{W}\ve{x}+\ve{n},
\end{equation}
where $\ve{d}$ is the CMB vector data, $\ve{n}$ is the noise vector, characterized by zero mean and covariance matrix
\begin{equation}
\m{N}=\big< \ve{n} \ve{n^{t}} \big>,
\end{equation}
and $\ve{x}$ is the vector of the unknowns. The CMB measurements are given by the angular power spectrum evaluated at the multipoles $\ell$, $\m{W}$ is the transfer function calculated numerically with a Boltzmann code such as CMBFAST (Seljak \& Zaldarriaga 1996), and $\ve{x}$ represents the initial power spectrum, as defined above. Cosmic variance is neglected here and $\m{N}$ is independent of $\ve{x}$.

Consider a Gaussian likelihood,

\begin{equation}
\label{eq:like}
\mathcal{L}\propto \exp \left[ -\frac{1}{2}(\m{W}\ve{x}-\ve{d})^{t}\m{N}^{-1}(\m{W}\ve{x}-\ve{d}) \right]
\equiv \exp \left[ -\frac{1}{2}S' \right],
\end{equation}
where we are ignoring an overall factor that does not depend on the model and Eq.(\ref{eq:like}) is implicitly defining $S'$. Our purpose is to find an estimate for the vector $\ve{x}$, given the data, their covariance matrix and the window function.

The solution to the minimization problem
\begin{equation}
{\partial \over \partial \ve{x} }  \left\{ S'[\ve{x}] \right\} = 0,
%\min\left\{ (\m{W}\ve{x}-\ve{d})^{t}\m{N}^{-1}(\m{W}\ve{x}-\ve{d}) \right\},
\end{equation}
maximizes the likelihood and provides a minimum variance estimator that is unbiased. The solution is found to be
\begin{equation}
\label{eq:xbar}
\overline{\ve{x}}=\left[ \m{W}^{t} \m{N}^{-1} \m{W} \right]^{-1} \m{W}^{t} \m{N}^{-1} \ve{d},
\end{equation}
and its covariance matrix is given by
\begin{equation}
\label{eq:Sigma}
\Sigma =\left[ \m{W}^{t} \m{N}^{-1} \m{W} \right]^{-1}.
\end{equation}
However in this context
it is not possible to compute the inverse matrix defined above, essentially because the transfer function $\m{W}$ does not represent a one-to-one relation connecting the data space to the solution space, since the projection on the last scattering surface and smearing effects (including gravitational lensing, that is however not considered in the present work) destroy part of the information (Tegmark \& Zaldarriaga 2002). Note that the formal solution provided by Eqs.(\ref{eq:xbar}) and (\ref{eq:Sigma}) is valid only under the assumption that $\m{N}$ is independent of $\ve{x}$.

Clearly to solve the problem, some external information must be injected in 
the form of a prior. In order to do this in a meaningful way, the problem must be rephrased in Bayesian terms. 
We are interested in evaluating the probability $P(\ve{x}|\ve{d})$ of having a primordial power spectrum given the data. Bayes theorem states that:
\begin{equation}
P(\ve{x}|\ve{d})=\frac{P(\ve{d}|\ve{x})P(\ve{x})}{P(\ve{d})},
\end{equation} 
where $P(\ve{d}|\ve{x})$ is the likelihood of the data given a model for the power spectrum, $P(\ve{x})$ is the prior assumed on the power spectrum and $P(\ve{d})$ is the probability of having the data. In what follows we will fix the late-time cosmological parameters to the best fit values found by \cite{tegpar} when they assumed a featurless power law for the initial power spectrum.

We will consider a prior that favours smooth solutions and a useful choice is to adopt a prior of the form
%\begin{equation}
%P(\ve{x})\propto e^{- \epsilon \ve{x}^{t}\m{L}^{t}\m{L}\ve{x}},
%\end{equation}
\begin{equation}
\label{eq:reg}
P(\ve{x})\propto \exp \left( -\epsilon \ve{x}^{t}\m{L}^{t}\m{L}\ve{x} \right)
\equiv  \exp \left( -\epsilon R \right) , 
\end{equation}
with $\m{L}$ representing an approximation to the first derivative operator and $\epsilon$ is a constant parameter. 
Eq.(\ref{eq:reg}) defines $R$ as the regularizing operator whose strength depends on $\epsilon$.
In other words, the logarithm of the prior, namely the regularization  
function,  approximates a penalty term that corresponds to the squared norm of the first derivative of the solution. 
After neglecting the probability of the data, that does not depend on the initial power spectrum, the minimization problem generalizes to   
\begin{equation}
{\partial \over \partial \ve{x} }  \left\{ S'[\ve{x}] + \epsilon R[\ve{x}] \right\} = 0.
%\min \left\{ (\m{W}\ve{x}-\ve{d})^{t}\m{N}^{-1}(\m{W}\ve{x}-\ve{d})+\epsilon \ve{x}^{t}\m{L}^{t}\m{L}\ve{x} \right\}.
\end{equation}
The parameter $\epsilon$ weighs the amount of smoothing that the solution is
requested to have. In other words, it is possible to find a solution, provided it is smoothed to a certain degree. 
Under the assumption that the covariance matrix $N$ does not depend on the theoretical power spectrum, the solution can be written as
\begin{equation} \label{sol}
\overline{\ve{x}}=\left[ \m{W}^{t}\m{N}^{-1}\m{W}  +\epsilon \m{L}^{t}\m{L} \right]^{-1} \m{W}^{t}\m{N}^{-1}\ve{d},
\end{equation}
with the now computable covariance matrix
\begin{equation}
\label{eq:sig}
\Sigma=\left[ \m{W}^{t}\m{N}^{-1}\m{W} +\epsilon \m{L}^{t}\m{L} \right]^{-1}. 
\end{equation}
The dependence of the solution on the smoothness parameter $\epsilon$ appears clearly in the previous equations. It is useful to view some asymptotic limits of the solution. When the parameter $\epsilon$ tends to zero we are back to the first case of no prior and the solution is going to be impossible to find for the reasons exposed above. If $\epsilon$ is taken very large and the smoothing term dominates over the first term, the solution will be in the form that minimizes the first derivative, that is, a constant vector.

Clearly, a compromise is needed when the parameter $\epsilon$ has to be chosen. A reasonable choice is to select an $\epsilon$ such that
\begin{equation}
S'[\overline{\ve{x}}] =N_d
%(\m{W}\overline{\ve{x}}-\ve{d})^{t}\m{N}^{-1}(\m{W}\overline{\ve{x}}-\ve{d})=N_{d},
\end{equation}
with $N_{d}$ representing the number of data points. In essence, this was the
method adopted in our first reconstruction paper, ~\cite{dtv}, where, for
high enough multipoles, cosmic variance could be considered negligible 
with respect to experimental noise and a Gaussian shape for the likelihood of
the data given a model that  was a safe choice, as will also be confirmed below.

Clearly the algorithm can 
now be cast as a classical Lagrange multiplier problem. Namely the equations to be solved are
\begin{equation}
\label{eq:lagrange1}
S'[\ve{x}] =N_d
\end{equation}
and 
\begin{equation}
\label{eq:lagrange2}
{\partial \over \partial \ve{x} }  \left\{ S'[\ve{x}] + \epsilon R[\ve{x}] \right\} = 0.
\end{equation}

\subsection{Extension to Large Scales}
\label{subsec:ext}

Since we are interested in an estimate of the power spectrum at large scales, corresponding to large angle CMB data, the simple Gaussian likelihood of the previous section can no longer be used. The first non-Gaussian modifications to get a reliable fit for the CMB likelihood and the need for them were proposed in Bond, Jaffe, Knox (1998 and 2000). 
For the WMAP experiment, the likelihood is approximated in detail by the product of a lognormal and a Gaussian distributions (Verde et al. 2003): 
\setlength\arraycolsep{2pt}
\begin{eqnarray} \label{likcomp}
-2\ln \mathcal{L} & = & \frac{2}{3}(\ve{z}^{th}-\ve{z}^{d})^{t}\m{Q}^{-1}(\ve{z}^{th}-\ve{z}^{d})+
\nonumber\\
& &\frac{1}{3}(\m{W}\ve{x}-\ve{d})^{t}\m{N}^{-1}(\m{W}\ve{x}-\ve{d}),
\end{eqnarray}
with $\ve{z}^{th}=\ln (\m{W}\ve{x}+\ve{\mathcal{N}})$ and $\ve{z}^{d}=\ln (\ve{d}+\ve{\mathcal{N}})$.
Here 
\begin{equation} \label{qmatr}
Q^{-1}_{\ell\ell'}= (\m{W}\ve{x} + \ve{\mathcal{N}})_\ell N^{-1}_{\ell\ell'} (\m{W}\ve{x} + \ve{\mathcal{N}})_{\ell'}. 
\end{equation} 
The noise vector $\ve{\mathcal{N}}$ was computed by the WMAP team (Verde et al. (2003)) through calibration with Monte Carlo simulations.
It is important to note that the matrix $\m{N}$ depends on the angular power spectrum $\m{W}\ve{x}$, due to cosmic variance and in addition the galaxy cut provokes some correlation due to mode mixing. 

Since the primordial power spectrum reconstruction on large scales is
very sensitive to the quadrupole likelihood function, it is relevant to be 
sure that foreground contamination has been properly taken into account. There have been various post-WMAP papers in which alternative methods to clean foregrounds
have been presented (e.g. in Tegmark et al. (2003), Efstathiou (2004) and Slosar et al. (2004)). We will discuss later the possible effects of alternative foreground removal methods.  

Even if in this work we chose to use the Verde et al.(2003) likelihhood, our method can be generalised provided an analytical fit to the likelihood is given.

Application of Bayes theorem and the requirement that the estimated power
 spectrum should correspond to the maximum of the \emph{a posteriori}
 probability reduces
 the minimization problem to
\begin{equation}
\label{eq:SeR}
{\partial \over  \partial \ve{x}}  \left\{ S  + \epsilon R \right\} = 0,
\end{equation}
where
\begin{eqnarray}
S &=& \frac{2}{3}(\ve{z}^{th}-\ve{z}^{d})^{t}\m{Q}^{-1}(\ve{z}^{th}-\ve{z}^{d})+  \nonumber\\
& & \frac{1}{3}(\m{W}\ve{x}-\ve{d})^{t}\m{N}^{-1}(\m{W}\ve{x}-\ve{d}).
%+\epsilon \ve{x}^{t}\m{L}^{t}\m{L}\ve{x},
\end{eqnarray}
under the constraint of 
\begin{equation}
\label{eq:SNd}
S[\ve{x}] =N_d.
\end{equation}
Here the covariance matrix $\m{N}$ has an explicit dependence on $\ve{x}$ to account for the cosmic variance and $N_{d}$ equal to 899, the number of WMAP temperature data points.

The minimization of $\left(S  + \epsilon R \right)$ cannot be solved analytically, due to the non-Gaussianity of the likelihood function, therefore a numerical method should be used; we found the iterative Newton-Raphson method suitable for our purpose.  

The $m+1$-th iteration of the method is decribed by:
\begin{equation}
x_{i}^{m+1}=x_{i}^{m} - {1 \over 2} \sum_{j} H^{-1}_{ij} \frac{\partial \left( S+\epsilon R \right)}{\partial x_{j}}\bigg|_{\ve{x}^{m}},
\end{equation}
in which $x^{m}_{i}$ stands for the solution $i$-th component relative to the $m$-th iteration. 
Here we have the following partial derivative:

\setlength\arraycolsep{2pt}
\begin{eqnarray}
\lefteqn{ \frac {\partial \left(S+ \epsilon R \right)}{\partial x_{i}}\bigg|_{\ve{x}^{m}} =}
\nonumber\\
& & \frac{2}{3} \sum_{\ell \ell'} W_{\ell i} N^{-1}_{\ell \ell'} (\m{W}\ve{x}^{m}+\ve{\mathcal{N}})_{\ell'} ( \ve{z}^{th}- \ve{z}^{d} )_{\ell'}+
\nonumber\\
&  & \frac{2}{3} \sum_{\ell \ell'} ( \ve{z}^{th}- \ve{z}^{d} )_{\ell} W_{\ell i} N^{-1}_{\ell \ell'} (\m{W}\ve{x}^{m}+\ve{\mathcal{N}})_{\ell'} ( \ve{z}^{th}- \ve{z}^{d} )_{\ell'}+
\nonumber\\
&  & \frac{1}{3} \sum_{\ell \ell'} W_{\ell i} N^{-1}_{\ell \ell'}(\m{W}\ve{x}^{m}-\ve{d})_{\ell'} + \epsilon (\m{L}^{t}\m{L}\ve{x}^{m})_{i}-
\nonumber\\
&  & \frac{1}{6} \sum_{\ell} (\m{W}\ve{x}^{m}-\ve{d})^{2}_{\ell} (2\ell+1) f^{2}_{\ell} W_{\ell i}/(\m{W}\ve{x}^{m}+\ve{\mathcal{N}})^{3}_{\ell}-
\nonumber\\
&  & \frac{1}{3} \sum_{\ell}  ( \ve{z}^{th}- \ve{z}^{d} )^{2}_{\ell} (2\ell+1) f^{2}_{\ell} W_{\ell i}/(\m{W}\ve{x}^{m}+\ve{\mathcal{N}})_{\ell}.
\end{eqnarray}
It is intended that in the last equation $\ve{z}^{th}=\ln(\m{W}\ve{x}^{m}+\ve{\mathcal{N}})$ at each iteration. Furthermore the last two terms were found taking into account the derivative of the covariance matrix $\m{N}^{-1}$, assumed, just for these terms, to be diagonal and with diagonal elements given by:
\begin{equation}
N^{-1}_{\ell \ell}=\frac{(2\ell+1)}{2} \frac{f^{2}_{\ell}}{(\m{W}\ve{x}+\ve{\mathcal{N}})^{2}_{\ell}},
\end{equation}
in which $f_{\ell}$ is the effective fraction of the sky computed by the WMAP team (Verde et al. (2003)). 

Furthermore we have approximated the Hessian matrix with:
\begin{equation}
H_{ij}={1 \over 2} \frac {\partial^{2} \left( S+ \epsilon R \right) }{\partial x_{i} \partial x_{j}}\bigg|_{\ve{x}^{m}}\simeq \left( \m{W}^{t}\m{N}^{-1}\m{W}+\epsilon \m{L}^{t}\m{L} \right)_{ij},
\end{equation}
where we neglected difference terms that do not substantially alter the final result and may induce numerical instabilities. At the $(m+1)$-th iteration step the covariance matrix is evaluated at $\m{W}\ve{x}^m$. Eq.(\ref{eq:SeR}) is solved for an assumed value of $\epsilon$, upon convergence the value of $\epsilon$ is updated and the whole procedure is repeated until the 
two equations are both satisfied.

\subsection{Two types of errors}

With the iterative procedure we can identify the maximum of the a posteriori
distribution of the power spectrum. Given the maximum 
and considering small deviations around the maximum, then the distribution function can be approximated as Gaussian and can be fully defined by the error covariance matrix. 
To estimate the statistical significance of the estimated power spectrum we define two types of errors.

According to the Bayesian method, the covariance matrix that describes the scatter of the  true $\ve{x}$ around its estimator, $\bar{\ve{x}}$ is given by: 
\begin{equation} \label{errorsprior}
\Sigma_{I}= \big< (\ve{x}-\bar{\ve{x}})(\ve{x}-\bar{\ve{x}})^{t} \big>=\left[ \m{W}^{t}\m{N}^{-1}\m{W} +\epsilon \m{L}^{t}\m{L} \right]^{-1}.
\end{equation}
Note that $\Sigma_{I}$ provides an estimate of the deviation of the true $\ve{x}$ from its estimator, and it is referred to as an error of type I.
   
Alternatively, one can ask for another measure of the scatter, this time of frequentist nature. Let us assume for a moment that the likelihood of the data is Gaussian and that the estimator can be expressed as:
\begin{equation}
\overline{\ve{x}}=\m{M}\ve{d},
\end{equation}
with
\begin{equation}
\m{M}=\left[ \m{W}^{t}\m{N}^{-1}\m{W} +\epsilon \m{L}^{t}\m{L} \right]^{-1} \m{W}^{t}\m{N}^{-1}.
\end{equation} 
Suppose that an ensemble of observers is measuring the same realization of the primordial power spectrum $\ve{x}$, and that each observer measures and estimates it in the same way. 
It is clear that the ensemble averaged value of the estimated $\ve{x}$ is given by:
\begin{equation}
\ve{x}_{ensemble}=\big< \bar{\ve{x}} \big> =\m{M}\m{W}\ve{x}.
\end{equation}
The variance of the scatter of the estimated $\ve{x}$ from its ensemble average is defined here as an error of type II, and is given by:
\begin{equation}
\Sigma_{II}=\big< (\bar{\ve{x}}-\ve{x}_{ensemble})(\bar{\ve{x}}-\ve{x}_{ensemble})^{t} \big>=\m{M}\m{N}\m{M}^{t}.
\end{equation}
In other words, $\Sigma_{II}$ gives the scatter in the estimator of $\ve{x}$
 around its mean value for a given realization. This corresponds to using
 the estimator of $\ve{x}$ as a surrogate for the true model and to employing
 synthetic data built around it to study the distribution of the errors.

Suppose that the estimation method used here, represented by the matrix $\m{M}$, does not introduce any bias in $\ve{x}$, then it follows that $\Sigma_{II}$ provides a measure of how much $\bar{\ve{x}}$ deviates from the true value. It will be shown that the bias is almost negligible for the very smooth concordance theoretical spectra and that the criterion in Eq.(\ref{eq:SNd}) tends to oversmooth the predicted signal. Consequently $\Sigma_{II}$ presents a lower limit on the scatter of the estimated $\ve{x}$ from its theoretically expected shape. 

In the next section,  Monte Carlo simulations generated around the
concordance initial spectrum are used to show that in the general 
non-Gaussian WMAP likelihood case,  practically no bias is present and that
the covariance matrix computed from the \emph{a posteriori} distribution of
the deconvolved samples is almost identical to $\Sigma_{II}$, as expected
from error propagation. Furthermore the distribution  very closely follows a Gaussian. 
\subsection{Null Hypothesis Testing}

The smoothing prior necessarily induces  a bias, that is caused by the matrix
($\m{M}\m{W}$) being different with  respect to unity. In other words, the deconvolved power spectrum, even in absence of noise, is inevitably going to be a smoothed version of the true spectrum.

However the actual concordance model power spectra are predicted by the simplest inflation scenarios and are given by very smooth shapes, namely a power law with an exponent $n_{s}$ very close to 1, that result in a very flat combination $k^{n_{s}-1}$. Therefore a smoothing prior seems reasonable both on theoretical grounds (since it is in line with inflationary predictions) and for practical purposes (since it offers an effective way to extract informations given the limitations posed by the width of the transfer function kernel). Remarkably, for extremely flat configurations the bias caused by the smoothing prior adopted in our method can be considered nearly negligible. Indeed this is what we have found in the Monte Carlo tests explained in the next section, where we considered two power law spectra with $n_{s}=1$ and 0.97. The fact that the method has nearly negligible bias for configurations close to the concordance model has led us to consider an extremely effective Null Hypothesis Test that consists of the following steps:
\begin{itemize}
\item the Null Hypothesis is chosen to be given, in turn, by the two initial spectra with $n_{s}=1$ and 0.97;
\item the CMB angular power spectra are computed from the initial spectra and from them random deviates are drawn to get synthetic data;
\item each of the Monte Carlo realizations are used as starting points for the reconstruction method and the deconvolved solutions are utilised to sample the a posteriori distribution of the initial spectrum around the Null Hypothesis;
\item the reconstruction from WMAP data is compared with the a posteriori distribution and if strong deviations are found, then the Null Test is considered to have failed.
\end{itemize}

We remark that if the real primordial spectrum has deviations from a smooth shape, the bias introduced by the method will tend to conservatively underestimate them.

It is useful to mention that the errors of type II built around the Null Hypothesis spectra are expected to be equivalent to the errors that result from the scatter of a large number of Monte Carlo realizations, and indeed this is what we have observed.
As specified in the next Section, to quantify the probability of how likely the Null Hypothesis is followed in pre-defined wavenumber ranges, we have integrated over the distribution described by the Monte Carlo samples. 
Clearly, all these considerations require that the important systematics have been removed from the data; a discussion on this delicate issue is presented in the next Section.

\section{Results}
\label{sec:results}

Since the numerical computations are rather intensive, we fixed the late-time cosmological parameters to the best-fit values found combining WMAP temperature and SDSS data~\cite{tegpar}: for the baryonic and matter fractions, $\Omega_{b}h^{2}=0.0231$ and $\Omega_{m}h^{2}=0.149$; the Hubble constant in units of $100 \, \mathrm{Km \, s^{-1} \, Mpc^{-1}}$, $h=0.685$. The optical depth was fixed to $\tau=0.166$, 
as suggested by the WMAP temperature and cross correlated temperature-polarization data, when the power spectrum is described by a power law with spectral index close to unity (Spergel et al. 2003). The range over which the transfer functions are integrated is typically between $10^{-5}$ and $0.1 \, \mathrm{Mpc^{-1}}$, discretised in approximately 500 bins.
This choice corresponds approximatively to a marginalization if the precise question  we are answering is whether there are deviations from a power law spectrum with spectral index close to
$n_{s}=0.97$, that was found in the best fit of \cite{tegpar}, while assuming a power spectrum defined as a power law with the spectral index as a free parameter. 
In other words, the method can be thought of as a consistency test for the concordance model with a power law power spectrum, since by fixing the cosmological parameters in this way, we should expect to find a power spectrum at least compatible with a power law of $k^{n_{s}}$ with $n_{s}=0.97$.
Under these conditions the prior will tend to push the reconstruction towards the above 
power law, if it represents the true underlying signal. While if departures
from featureless initial conditions are found with enough statistical
significance, they may be due to a real signal or just to residual systematics still present in the data. We have also implemented the analysis with a spectral index $n_{s}=1$ to look for deviations from a scale free spectrum.

As anticipated, an additional safety net is offered by the method: if large
enough deviations truly exist, the estimate obtained with the smoothing 
criterion in Eq.(\ref{eq:SNd}) will in general tend to underestimate the signal by furnishing an oversmoothed solution. This is a clear result of the prior bias that we have verified with a series of tests and is also deducible from a computation of the effective number of degrees of freedom caused by smoothing. These are defined, as in \cite{dtv}, by the trace of the matrix (\m{W}\m{M}) in analogy with ordinary regression and amount to about 52 for the WMAP reconstruction. Eq.(\ref{eq:SNd}) should therefore be considered a rather conservative choice. 

To verify the method we performed Monte Carlo tests. Assuming a power law with spectral index $n_{s}=0.97$ and with $n_{s}=1$ as an input, we calculated the corresponding CMB angular power spectrum and we created random realizations around it. Neglecting correlations, we sampled the probability distribution derived from the noise model proposed in Knox (1995):
\begin{equation}
P(d_{\ell})=\frac{n}{C_{\ell}+\mathcal{N}_{\ell}} \frac{V^{(n-2)/2}e^{-V/2}}{2^{n/2}\Gamma(n/2)},
\end{equation}
   where $C_{\ell}$ is the theoretical spectrum, $n=f_{\ell}^{2}(2\ell+1)$ and
\begin{equation}
V=n\frac{d_\ell+\mathcal{N}_{\ell}}{C_\ell+\mathcal{N}_{\ell}}.
\end{equation}
This distribution has the mean equal to the theoretical spectrum, $\big< d_\ell \big>=C_{\ell}$, and variance
\begin{equation}
\big<(d_\ell-C_\ell)^{2}\big>=\frac{2}{n}(C_\ell+\mathcal{N}_{\ell})^{2}.
\end{equation}
In other words we have incorporated in the noise model the fit of the diagonal of the Fisher matrix $\m{N}^{-1}$ obtained by the WMAP team (Verde et al. 2003) through monte carlo realizations of the sky that included noise, beam and cut sky effects.

\begin{figure}
\includegraphics[totalheight=0.19\textheight,viewport=110 150 200 200,clip]{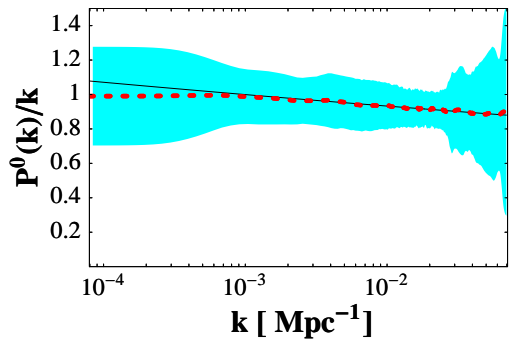}
\caption{\label{fig.1} Results from 1000 Monte Carlo realizations drawn from a power law power spectrum, given by $k^{n_{s}}$ where $n_{s}=0.97$, with the concordance WMAP-SDSS cosmological parameters of Tegmark et al.(2004). 
The thin smooth line in the centre is the exact initial power law and the dashed curve is the averaged reconstruction. Also shown are the standard deviation bands calculated from the scatter of the realizations around the mean. Practically no bias appears in the reconstruction, except at very large scales, where the prior starts to dominate, and however the bias is considerably smaller than the error estimation. The plot is in units of $2.95\times 10^{-9}$.}
\end{figure}

\begin{figure}
\includegraphics[totalheight=0.19\textheight,viewport=110 150 200 200,clip]{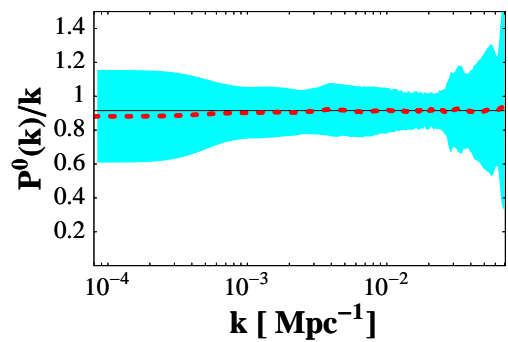}
\caption{\label{fig.1.0} Results from 1000 Monte Carlo realizations drawn from a power law power spectrum, given by $k^{n_{s}}$ where $n_{s}=1$, with the concordance WMAP-SDSS cosmological parameters of Tegmark et al.(2004). The thin smooth line in the centre is the exact initial power law and the dashed curve is the averaged reconstruction. Also shown are the standard deviation bands calculated from the scatter of the realizations around the mean. Practically no bias appears in the reconstruction, except at very large scales, where the prior starts to dominate, and however the bias is considerably smaller than the error estimation. The plot is in units of $2.95\times 10^{-9}$.}
\end{figure}

To speed up the computations, we considered just the diagonal elements of the inverse of the covariance matrix $\m{N}^{-1}$, and consequently of the inverse of the matrix $\m{Q}^{-1}$ in Eq.(\ref{qmatr}), and we created samples of the power separately at each multipole. We verified that the reconstructions operated with the full or diagonal matrices $\m{N}^{-1}$ and $\m{Q}^{-1}$ do not differ appreciably.  Fig.~\ref{fig.1} and Fig.~\ref{fig.1.0}  show the outcome of the test and it is relevant to notice that the bias that results from the reconstruction is negligible when compared with type I and even type II errors. This conclusion supports our consideration of the reliability of the less conservative type II errors, when utilised to detect the departure from the flat and concordance power law models. We also noted that type II errors are almost identical to the standard deviation uncertainties computed from the Monte Carlo samples; this provides a useful check for the method.

For all the reconstructions from the Monte Carlo realizations, we fix
the parameter $\epsilon$ to the value that results from deconvolving
the WMAP data according to Eq.(\ref{eq:SNd}). We have also checked
that when choosing $\epsilon$ following Eq.(\ref{eq:SNd}) for each
individual Monte Carlo sample built from an initial spectrum with
features added, the resulting mean reconstruction was a conservative
smoothed version of the starting one.

The analysis of the Monte Carlo realizations validates that the
algorithm used here indeed produces an almost unbiased estimator of power
spectrum, when very flat spectra are used as input. If the power spectrum has real features, then some bias will be inevitably present in the estimator, but it will act in order to conservatively underestimate the features.

We verified that the Monte Carlo reconstructions on all scales are
well described by a Gaussian distribution that follows type II
errors.

\begin{figure}
%[!h]
\includegraphics[totalheight=0.2\textheight,viewport=110 160 200
210,clip=]{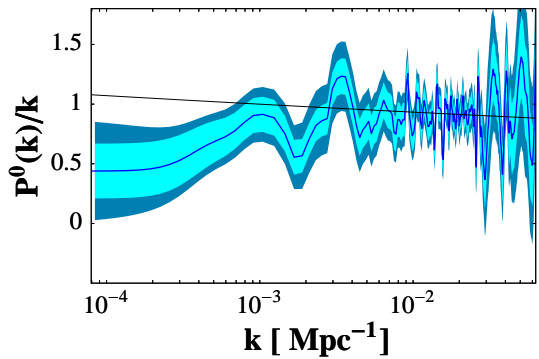}
\caption{\label{fig.3} Reconstructed initial power spectrum in units
of $2.95\times 10^{-9}$ from WMAP data with the concordance model
cosmological parameters. The irregular middle curve represent the mean
and the surrounding curves that delimit the light shaded band are
obtained by summing and subtracting the square root of the diagonal
elements of the covariance matrix for type II errors. The borders of
the larger dark shaded region are given by summing and subtracting the
square root of the diagonal elements of the covariance matrix for type
I errors.  The smooth line is a reference power law power spectrum
given by $k^{n_{s}}$, where $n_{s}= 0.97$.}

\end{figure}

\begin{figure}
\includegraphics[totalheight=0.19\textheight,viewport=110 150 200
200,clip]{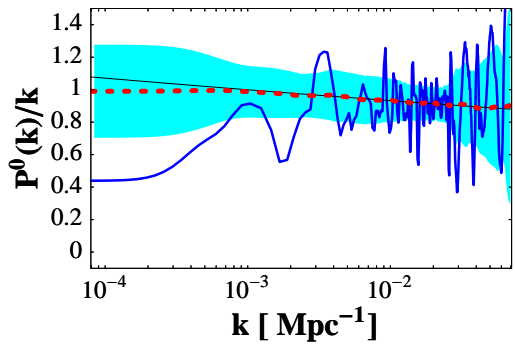}
%\includegraphics{fig.mc.ngauss.1.ps}
%\begin{figure}[!h]
%\includegraphics[totalheight=0.2\textheight,viewport=110 160 200
%210,clip=]{fig.mc.ngauss.1.ps}
\caption{\label{fig.3.0} Results from 1000 Monte Carlo realizations
drawn from a power law power spectrum, given by $k^{n_{s}}$ where
$n_{s}=0.97$, with the concordance WMAP-SDSS cosmological parameters
of Tegmark et al.(2004). The thin smooth line in the centre is the
exact initial power law and the dashed curve is the averaged
reconstruction. Also shown are the standard deviation bands calculated
from the scatter of the realizations around the mean. The thick
continuous line, the reconstruction from WMAP data, is also
plotted to show   how likely it is that  the deconvolution could
represent a random deviate of the power law fiducial model, assumed to
be our Null Hypothesis. The plot is in units of $2.95\times 10^{-9}$.}
\end{figure}

In Fig.~\ref{fig.3}  our most important result is plotted, namely the
estimated power spectrum from the WMAP temperature data (Bennett et
al. 2003) at large scales together with the two type of errors.  In
Fig.~\ref{fig.3.0} we show the WMAP reconstruction compared with the
Monte Carlo results described previously for the power law spectrum
case. The figure can provide intuition of how unlikely in some wave
number ranges the reconstruction could resemble a statistical random
deviation from the fiducial spectrum, and depicts a graphical version
of the Null Hypothesis Test described in the previous Section.

Assuming the flat spectrum to define  the fiducial
initial conditions, we find evidence for the three following potential
deviations from a power spectrum without features considering type I
(type II) errors: a cutoff for the largest scales $0.0001<k<0.001 \,
\mathrm{Mpc^{-1}}$ at 1 (2) sigma c.l.; a dip at $0.001<k<0.003 \,
\mathrm{Mpc^{-1}}$ at 1.5 (3) sigma c.l.; a bump a $0.003<k<0.005 \,
\mathrm{Mpc^{-1}}$ at 1 (2) sigma c.l..  While with the power law as
the fiducial spectrum, we find, considering type I (type II) errors: a
cutoff at $0.0001<k<0.001 \, \mathrm{Mpc^{-1}}$ at 1.5 (2.5) sigma c.l.; a
dip at $0.001<k<0.003 \, \mathrm{Mpc^{-1}}$ at 1.5 (3) sigma c.l.; a
bump a $0.003<k<0.005 \, \mathrm{Mpc^{-1}}$ at 1 (2) sigma c.l..

We now provide another way to estimate the deviations, that is
practically equivalent to integrating over the distribution defined by
the errors of type II built around the Null Hypothesis spectra in
defined ranges in wavenumber space. Following \cite{kogo1}, we define
a measure of the distance between a generic spectrum ($P^{g}$) and the
fiducial one ($P^{f}$) in the wavenumber range limited by $k_{1}$ and
$k_{2}$:
\begin{eqnarray}
D(k_{1},k_{2})&=&\int_{k_{1}}^{k_{2}} \mathrm{d}k
\left(\frac{P^{g}(k)}{k} - \frac{P^{f}(k)}{k}\right)^{2} \nonumber\\
&\simeq& \sum_{i=i_{1}}^{i_{2}} \Delta k_{i}
\left(\frac{P^{g}(k_{i})}{k_{i}} -
\frac{P^{f}(k_{i})}{k_{i}}\right)^{2}.
\end{eqnarray} 
We calculated the distance of all the Monte Carlo samples and the WMAP
reconstruction with respect to the fiducial spectrum and we counted how
many realizations had a distance larger than the one computed for the
WMAP data deconvolution.  In this way we could quantify the
probabilities of exceeding the observed deviations in WMAP data
respect to the fiducial models. We detected the following potential
departures from a scale free (spectral index $n_{s}=0.97$) initial
spectrum: a cut-off at $0.0001<k<0.001
\, \mathrm{Mpc^{-1}}$ at 79.5\% (92\%), a dip at $0.001<k<0.003 \,
\mathrm{Mpc^{-1}}$ at 87.2\% (98\%) and a bump at
$0.003<k<0.004 \, \mathrm{Mpc^{-1}}$ at 90.3\% (55.5\%). 

We add that the prior, similarly to the Weiner filter (see e.g. \cite{zar}), helps in retrieving useful information at the crossover between the signal and noise dominated regimes, if combined with Monte Carlo testing. Furthermore methods such as ours offer the possibility to penetrate the informational barrier built up by cosmic variance. 

\begin{figure}
%[!h]
\includegraphics[totalheight=0.2\textheight,viewport=110 160 200
210,clip=]{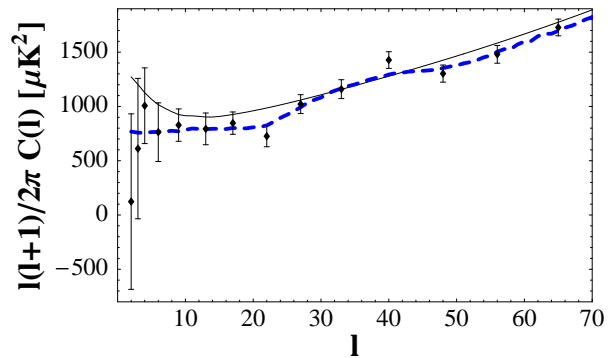}
\caption{\label{fig.4} The thick dashed smooth line represents the reconstructed power spectrum from WMAP data convolved back to multipole space, while the power law with spectral index $n_{s}=0.97$, best fit of the $\Lambda$CDM cosmology, is plotted as a continuous thick curve. The WMAP data are also shown as binned data points, with the error bars computed assuming the power law spectrum.} 
\end{figure}

We plot in Fig.~\ref{fig.4} the WMAP data together with the convolved estimated power spectrum in multipole space to illustrate the effect of smoothing. 
That interesting features might have been present in the power spectrum could have already been guessed from  an inspection of the WMAP data confronted with a power law power spectrum $\Lambda$CDM prediction. However how the information (including the experimental uncertainties) in multipole space was going to be transferred in detail to wave number space could not have been predicted reliably without an analysis that had taken properly into account the complicated transfer functions. Furthermore, the processed information is now in a format that permits a direct comparison with the theoretical predictions from inflation.

\begin{figure}
%[!t]
\includegraphics[totalheight=0.2\textheight,viewport=110 160 200
210,clip=]{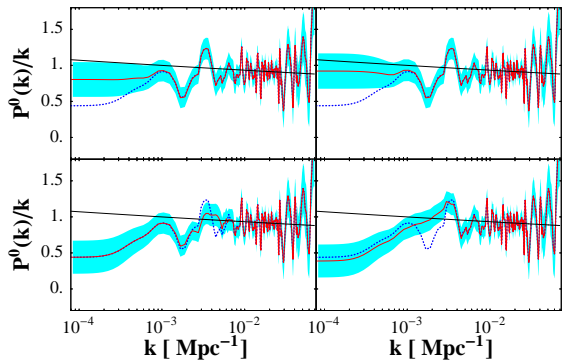}
\caption{\label{fig.5} The figure shows the effects of correlation and sheds light on the multipole space origin of the features detected. To serve such a purpose, we replaced some WMAP data points with the values of prediction of the $\Lambda$CDM concordance model with a power law power spectrum given by $k^{n_{s}}$, where $n_{s}=0.97$. The dotted curve is the WMAP reconstruction and the error bands are derived from type II errors (see text). Going from top left and clockwise the power was modified at the following multipoles: $\ell=2$; $\ell=2,3$; $15\leq \ell \leq 26$; $26 \leq \ell \leq 56$.} 
\end{figure}

Even though  we compute the full covariance matrices, in the plots we show the errors  given by the square root of the inverse diagonal elements of the relative covariance matrix for illustrational purposes. However it is important to assess if the deviations detected and their magnitude take their origin from error correlation. We implemented some tests to check this relevant issue. In turn, we replaced the WMAP multipoles that correspond to the scales of the three detected features with the concordance model predictions without any noise. From the reconstructions in Fig.~\ref{fig.5}, we deduce that the evidence for the features is not the product of correlation, since their significance is only slightly worsened. It is interesting to note that, as expected, the low values of the dipole and the quadrupole are the principal cause for the observed cutoff. Furthermore the wiggles in the WMAP data around the multipoles of 20 and 40 show up clearly as the bump and the dip in our reconstruction.

\begin{figure}
%[!h]
\includegraphics[totalheight=0.2\textheight,viewport=110 160 200
210,clip=]{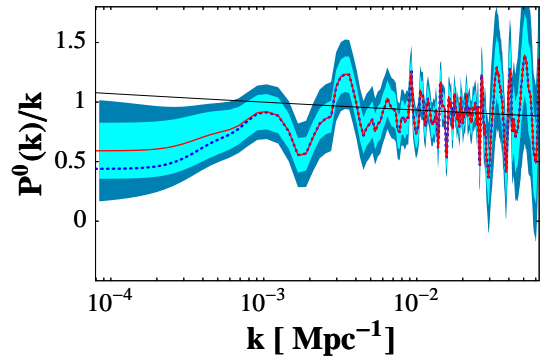}
\caption{\label{fig.6} The result of considering quadrupole and
octopole mean values raised to the Tegmark et al. (2003) all-sky
foreground subtracted map determination. The smooth curve represents a
power-law power spectrum given by $k^{n_{s}}$, where $n_{s}=0.97$,
while the dotted line is the deconvolution from the original WMAP
data. The continuous curve is the reconstruction with the new data,
the error bands come from the type I (darker) and type II (lighter)
errors. The smoothing parameter $\epsilon$ was fixed according to
Eq.(\ref{eq:SNd}) using the new data set.}
\end{figure}

We show Fig.~\ref{fig.6} how the reconstruction changes if we adopt
the analysis of \cite{tegmap}, based on an all-sky foreground cleaned
map derived from WMAP data. For representative purposes, we utilise the
WMAP data likelihood with the mean values of the quadrupole and
octopole lifted to respectively 202 and 870 $\mu\mathrm{K}^{2}$,
rather than using the original WMAP values of 123 and 612
$\mu\mathrm{K}^{2}$. It can be seen that, with respect to the
concordance power law spectrum with spectral index $n_{s}=0.97$, a
cut-off detection is still present at about $2\sigma$ in frequentist
terms, while the Bayesian estimate finds the $\ve{x}$ of the Null
Hypothesis $1\sigma$ away from the estimate $\bar{\ve{x}}$. We report that some other of the post-WMAP papers on the estimation of the quadrupole point towards a somewhat larger value respect to the original WMAP one (e.g. Efstathiou 2004 and Slosar et al. 2004). For example Slosar et al. (2004) find a broad quadrupole likelihood function with a tail that favours greater values, as a result of having marginalized over the foreground templates. This differs from the WMAP analysis that consisted in fitting and subtracting out the templates from the data. Clearly both the estimated mean and errors of the reconstructed initial power spectrum would be affected by the modification of the CMB dataset.

\begin{figure}
%[!h]
\includegraphics[totalheight=0.2\textheight,viewport=110 160 200
210,clip=]{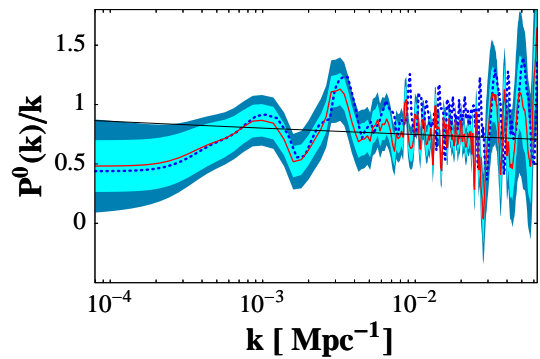}
\caption{\label{fig.7} We show the reconstruction for the
cosmological late-time parameters given in Spergel et al. (2003),
where an optical depth $\tau=0.1$ is considered. The smooth curve
represents a power law power spectrum given by $k^{n_{s}}$, where
$n_{s}=0.97$, while the dotted line is the deconvolution from the
original WMAP data. The continuous curve is the reconstruction with
the new parameters, the error bands coming from the type I (darker) and
type II (lighter) errors. The smoothing parameter $\epsilon$ was fixed
according to Eq.(\ref{eq:SNd}), taking into account the new transfer
function. The smaller optical depth causes a reduction in the
statistical significance of the cutoff.}
\end{figure}

We also
present in Fig.~\ref{fig.7} a reconstruction with the best fit
parameters found in Spergel et al. (2003), when considering a power
law spectrum: $\Omega_{b}h^{2}=0.023$ and $\Omega_{m}h^{2}=0.13$; the
Hubble constant in units of $100 \, \mathrm{Km \, s^{-1} \,
Mpc^{-1}}$, $h=0.68$; the optical depth, $\tau=0.1$. The main effect
with respect to the model considered previously is a reduction of the
cut-off significance due to a smaller optical depth, as the  main effect
would be to reduce the power at intermediate and small scales while
leaving  the power unaltered at large scales. A conspiracy between a
low optical depth and enhanced power in the quadrupole and octopole
could in principle  considerably decrease the detection of a cut-off;
such measurements should therefore be followed closely in future
experiments.
\section{Discussion}
Given the temperature data from the WMAP experiment we have
reconstructed at high resolution the primordial power spectrum at
large scales by adopting an iterative smoothing algorithm, built to
deal with the non-gaussianity of the data errors caused mainly by
cosmic variance.  We have fixed the parameters to the best fit
late-time cosmological parameters found by \cite{tegpar} in order to
search for deviations from the concordance featurless power law power
spectrum, with spectral index $n_{s}$ approximately equal to 0.97. We also implemented the analysis with a scale free spectrum. The
horizon scale for the model we are considering corresponds to the
wavenumber $k_{h}=4.52\times 10^{-4} \, \mathrm{Mpc^{-1}}$. By
creating and reconstructing from Monte Carlo realizations of CMB data,
we have verified that our proposed estimator is effectively unbiased
if a scale free or power law with spectral index $n_{s}=0.97$ power
spectra are assumed. This has led us to the detection of the following
three deviations from a scale free (spectral index $n_{s}=0.97$)
initial spectrum: a cut-off at $0.0001<k<0.001
\, \mathrm{Mpc^{-1}}$ at 79.5\% (92\%), a dip at $0.001<k<0.003 \,
\mathrm{Mpc^{-1}}$ at 87.2\% (98\%) and a bump at
$0.003<k<0.004 \, \mathrm{Mpc^{-1}}$ at 90.3\% (55.5\%).

We have also derived using two alternative ways a measure of the local uncertainties in
the estimates of deviations and effective location of the initial
power spectrum, labeled by errors of type I and II. The frequentist
analysis finds the low $k$ cutoff of the estimated power spectrum to
be at about $2.5\sigma_{II}$ away from the $n_s=0.97$ model, while in
the Bayesian analysis the model is about $1.5\sigma_I$ away from the
estimated spectrum.

The convergence in the observed low power at large scales by the WMAP
and COBE DMR experiments is certainly interesting and merits
attention.  Since further high quality large-scale temperature and
polarization CMB data are expected in the near future, for example
from the next data releases of the WMAP team and the future Planck
mission, any method that could shed light on the nature of the
primordial perturbations will prove to be extremely useful for  improving our understanding
of  the mechanism that is responsible for the generation of cosmic
structure.
\section*{ACKNOWLEDGMENTS}
DTV would like to thank the Hebrew University, where part of this work has been carried out, for its hospitality. DTV also acknowledges a Scatcherd Scholarship from the University of Oxford. DTV thanks Chuck Bennett, Andrew Jaffe and Glenn Starkman for useful discussions.
This research has been partially supported by the ISF-143/02 and the Sheinhorn Foundation (YH). YH acknowledges a Leverhulme visiting Professorship (University of Oxford) and the Mercator Gastprofessur (Astrophysikalisches Institut Potsdam).

%\bibliography{prolarge}% Produces thebibliography via BibTeX.

\bsp

\label{lastpage}
    
\end{document}